\documentclass[superscriptaddress,aps,twocolumn]{revtex4-1}
\usepackage{bm}
\usepackage{float}
\usepackage{amssymb}
\usepackage{amsmath}
\usepackage{multirow}
\usepackage{lineno}
\usepackage{graphicx}
\usepackage{comment}
\usepackage{soul}
\usepackage[colorlinks=true,citecolor=blue,linkcolor=blue]{hyperref}
\usepackage[usenames]{color}
\usepackage{subfigure}

\begin{document}

\title{High-efficient harmonic vortex generation from a laser irradiated hollow-cone target}

\author{Ke Hu}
\affiliation{Tsung-Dao Lee Institute, Shanghai Jiao Tong University, Shanghai 200240, China}
\affiliation{Collaborative Innovation Center of IFSA (CICIFSA), Shanghai Jiao Tong University, Shanghai 200240, China}
\author{Longqing Yi}
\thanks{lqyi@sjtu.edu.cn}
\affiliation{Tsung-Dao Lee Institute, Shanghai Jiao Tong University, Shanghai 200240, China}
\affiliation{Collaborative Innovation Center of IFSA (CICIFSA), Shanghai Jiao Tong University, Shanghai 200240, China}
\affiliation{Key Laboratory for Laser Plasmas (Ministry of Education), School of Physics and Astronomy, Shanghai Jiao Tong University, Shanghai 200240, China}

\date{\today}

\begin{abstract}
It has been recently reported that ultraviolet harmonic vortices can be produced when a high-power circular-polarized laser pulse travels through a micro-scale waveguide.
However, the harmonic generation quenches typically after a few tens of microns of propagation, due to the build-up of electrostatic potential that suppresses the amplitude of the surface wave.
Here we propose to utilize a hollow-cone channel to overcome this obstacle. When traveling in a cone target, the laser intensity gradually increases, allowing the surface wave to maintain a high amplitude for a much longer distance. The harmonic vortices can be produced with very high efficiency. According to three-dimensional particle-in-cell simulations, the overall efficiency are boosted by almost one order of magnitude, reaching $>20\%$. It is also found that using the cone targets can mitigate efficiency decline due to pre-heating of plasma by laser prepulse. The proposed scheme paves the way to the development of powerful optical vortices sources in the extreme ultraviolet regime - an area of significant fundamental and applied physics potential.
\end{abstract}
\maketitle

Optical vortices refer to light beams that carry orbital angular momentum (OAM) and hence exhibit helical wave fronts \cite{Allen1992,Yao2011,Bliokh2015}.
The helical profile is described by a spatial phase term $\exp(-il\phi)$, where $\phi$ is the azimuthal angle, and the integer number $l$ is the topological charge, denoting that each photon carries an OAM equals to $l\hbar$. Such vortex beams can serve as a powerful tool in optical trapping \cite{Oneil2002}, optical communication \cite{Wang2012, Gibson2004, Willner2015}, quantum optics \cite{Leach2010,Malik2016} and biophotonics \cite{Willig2006}. In the high energy density physics, the high-power lasers carrying OAM provide an extra degree of freedom to control the laser-plasma interactions \cite{Onoda2004,Bliokh2008,Padgett2011,Vieira2018}.

In particular, high-frequency vortex beams are of great interest for probing and topologically controlling ultrafast physics processes \cite{Patchkovskii2012,Veenendaal2007}.
However, it is challenging to produce such lights with traditional spiral phase optics \cite{Biener2002,Sueda2004,Yao2011} because of its small wavelength. Therefore, high-harmonic generation (HHG) are usually relied on to produce optical vortices in the ultra-violet regime and beyond.
In particular, the relativistic oscillating mirror (ROM) \cite{Bulanov1994,Lichters1996,Baeva2006} is the most mature mechanism in the high energy density field, thus receives most attention. Two scenarios have been mostly considered: the first uses plasma holograms \cite{Leblanc2017} to allow the reflected harmonic beams gain OAM;
the second approach employs a relativistic Laguerre-Guassion (LG) pulse \cite{Zhang2015,Denoeud2017} or a circularly polarized (CP) laser \cite{Wang20191,Li2020,Zhang2021} irradiating a planar target, the orbital or spin angular momenta of the driver are then converted to the OAM of the harmonics by the helical electron oscillation on the mirror surface. However, in general the ROM mechanism is suppressed for a CP driving laser at normal incidence due to lack of oscillating terms in the laser ponderomotive force, it remains challenging for producing CP vortex beams with high-intensity.

\begin{figure*}[t]
\centering
\subfigure{\includegraphics[width=0.9\textwidth]{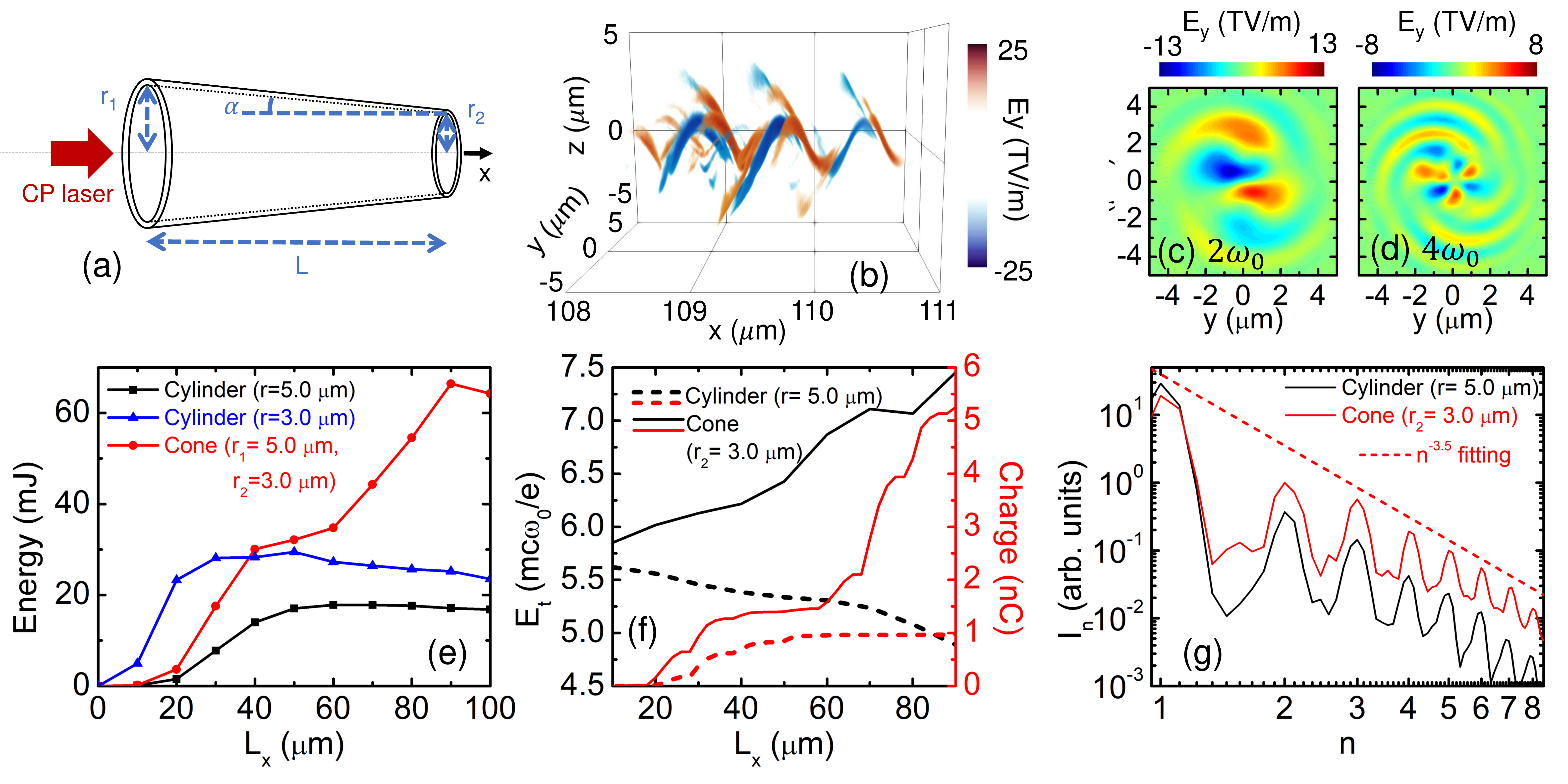}}
\caption{(a) Sketch of the proposed scheme and introduction of the target parameters.
(b) The high-frequency component of the electric field $E_y$ ($\omega>1.5\omega_0$) of the harmonic vortex when it just leaves the exit of the cone target.
The $E_y$ field of the (c) second- and (d) fourth-order harmonics are shown in the $y-z$ cross section, respectively.
(e) The harmonic energy plotted as a function of laser propagation distance $L_x$ for different targets.
%$r_2$, where $r_1=5.0\ \mu{\rm m}$ is fixed.
(f) Averaged amplitude of the transverse electric field $\bar{E_t}$ (black), and the charge of the accelerated electrons (red) versus laser propagation distance for the conical and cylindrical targets with $r_1=5\ \mu{\rm m}$ fixed.
(g) Spectra of the electromagnetic waves at the exit of the target for the two cases shown in (f).
    }
\end{figure*}

In our previous works, we proposed an alternative method to produce such advanced light sources by diffraction at relativistic intensities, namely the  ``relativistic oscillating window (ROW)" \cite{Yi2021}. It is later found that when using a micro-plasma waveguide, the ROW process continuously takes place as the laser traveling in the hollow channel. A self-phase-matching effect ensures the harmonic beams that generated at different longitudinal positions add up coherently, resulting in an ultra-intense vortex beam, the overall efficiency can reach $\sim 1\%$ \cite{Hu2022}. The main limitation on the HHG efficiency is a pre-mature electron over-extraction, meaning that too many electrons are extracted out of the waveguide wall by the driving laser field within a short distance. As a result, an electrostatic potential quickly builds up and suppresses the amplitude of the surface wave that are responsible for producing high-order harmonics.
Consequently, the HHG process typically quenches after the a few tens of microns of propagation, even the waveguide is much longer and the driving laser energy is still high.

In this letter, we propose to use high-power CP laser pulse interacting with a hollow-cone target to overcome this obstacles.
This kind of targets have been theoretically and experimentally proven to be a useful tool for
promoting the yield of thermal neutrons in the fast ignition research \cite{Kodama2001,Theobald2011},
%Besides, laser interactions with such hollow conical targets can
and are widely used in the acceleration of high-charge, high-energy ion bunches \cite{Chen2005,Zhu2022}, generation of attosecond pulses \cite{Lecz2016} and terahertz radiations \cite{Cai2022}.
By the means of three dimensional (3D) particle-in-cell (PIC) simulations, we demonstrate the overall efficiency for the high-harmonic vortex beam production can be boosted to $> 10\%$. This dramatic enhancement is mainly attributed to the slow focusing of the driving laser as it is guided through a cone channel,
%a large opening at the entrance mitigates electron over-extraction at beginning, while
the progressive laser intensification \cite{Sentoku2004,Budriga2020} manages to continuously overcome the established electrostatic barrier and maintain the surface wave amplitude at its maximum.
In addition, the utilizing of cone channels also relaxes the requirement for alinement and laser contrast, making the proposed scheme more accessible with current and upcoming laser systems.

We first present our scheme using a 3D PIC simulation with the EPOCH code \cite{Arber2015}.
A sketch of the laser-cone interaction setup, as well as an illustration the cone target parameters are shown in Fig. 1(a).
A CP driving laser pulse is tightly focused onto the entrance of a hollow cone target
along the $x$ axis from the left and travels to the right.
The simulation box has dimensions of $x\times y\times z=10\times 14\times 14\ \mu{\rm m}^3$ and is sampled by $1500\times 560\times 560$ cells, with four macroparticles for electrons and two for C$^{6+}$ per cell.
A moving window is used to improve computational efficiency, which follows the propagation of the driving pulse along the $x$ axis.
The laser field is $\mathbf{E}=(\mathbf{e_y}+i \mathbf{e_z})E_0 \text{exp}(-r^2/w_0^2) \text{exp}(-t^2/\tau_0^2)\text{exp}(ik_0x-i\omega_0t)$, where $\mathbf{e_y}$ ($\mathbf{e_z}$) are the unit vectors in $y(z)$ direction, $E_0$ is the laser amplitude, $\omega_0$ is the angular frequency, $k_0=2\pi/\lambda_0$ is the wavenumber, with $\lambda_0=1\ \mu{\rm m}$ the laser wavelength, $w_0=2.5\ \mu{\rm m}$ the laser spot size, and $\tau_0=6.79\ {\rm fs}$, corresponding to a temporal FWHM duration of $8.0\ {\rm fs}$.
The laser has a normalized amplitude $a_0=eE_0/mc\omega_0=16$, where $c$ is the speed of light, $m$ is the electron mass, and $e$ is the unit charge.
The cone target has a length of $L=100\ \mu{\rm m}$, an inner radius of $r_1=5.0\ \mu{\rm m}$ at the entrance.
The target has a uniform density of $n_0=30n_c$, where $n_c=\epsilon_0m\omega_0^2/e^2$ is the critical density. Here we use a sharp density ramp at the plasma boundary, the pre-plasma due to laser heating by the prepulse will be considered later.
We note that in this study we only consider small cone angles with $\alpha<5^\circ$,
as the focus spot of a high power drive laser is typically a few microns, if the cone angle is too large ($\alpha\gg5^\circ$), it only interacts with the rear of the cone channel.
For the representative case shown in Fig. 1, the radius at the exit is $r_2=3.0\ \mu{\rm m}$, meaning the cone angle is $\alpha=1.14^{\circ}$.

The harmonic vortices ($\omega/\omega_0>1.5$) generated from laser-cone interaction are illustrated by the color-coded electric field $E_y$ in Fig.~1(b). The helical phase front of the harmonic light arises from spin-orbit interaction facilitated by a chiral surface wave that co-propagate with driving laser pulse \cite{Hu2022}.
In Figs.~1(c-d), the field distributions of the second and fourth harmonics in the cross section perpendicular to the propagation axis are presented. One can see that
the topological charge $l$ and harmonic number $n = \omega/\omega_0$ satisfies a linear relationship $l = (n-1)\sigma$, where $\sigma = +1$ or $-1$ denotes the right- or left-handed circular polarization of the drive laser, respectively.
Therefore a helical structure of a light spring arises \cite{Pariente2015}, where the screw pitch equals the wavelength of the drive laser, and the peak electric field reaches 20 TV/m.
Such intense exotic  light is of interest for topological controlling of laser-plasma interaction, such as laser wakefield accelerators \cite{Vieira2018}.

The enhancement of HHG by a cone channel is shown in Fig.~1(e), where the energies of the high-order harmonics are plotted against laser propagation distance for both cone and cylindrical targets.
One can see that in the cylindrical channel cases ($r_2=r_1=3~\&~5\ \mu{\rm m}$), the harmonics generation saturate within $\sim50\ \mu \rm{m}$ of propagation. While in the cone channel, the harmonic beam energy continues to grow until the driving laser pulse reaches the exit, the total energy of the harmonic vortices is $63.5\ {\rm mJ}$, corresponding to a conversion efficiency of $11\%$.

The saturation of HHG energy is associated with the suppression of surface wave amplitude, which is responsible for harmonics generation and converting the SAM of fundamental light to the OAM of the harmonics \cite{Hu2022}.
As a high-power laser propagates in a plasma waveguide, it extracts electrons out of channel wall and accelerate them \cite{Yi2019,Hu2021}. The total charge of such electrons as shown by the red lines in Fig.~1(f).
By comparing with Fig.~1(e), one can see a similar trend between the accelerated electron charge and the harmonic energy for both cylindrical and conical targets.
This is because the electrons are extracted out via a surface-wave-breaking (SWB) effect. The increasing of electron charge thus indicates that the local surface wave amplitude, which is closely related the harmonic generation efficiency  \cite{Yi2021}, is at its maximum.

In a cylindrical waveguide, the SWB doesn't last long, because a electrostatic potential builds up as large amount of electrons are lost on the wall, which eventually suppresses the surface electron oscillation amplitude. In addition, the laser intensity decreases significantly with propagation distance (black dashed curve in Fig.~1(f)), thus leads to a premature quenching of the HHG process. On the other hand, the accelerated electron charge keep rising in a cone channel, meaning the surface wave amplitude is at its wave-breaking limit during the entire interaction time, so that the harmonics are continuously produced. This is mainly attributed to the progressively intensification of the drive laser pulse as it is focused by the cone target, as shown by the black solid curve in Fig.~1(f). The increasing laser intensity overcomes the established electrostatic potential, keeping the surface wave amplitude from decaying.

The effect is further confirmed by the harmonic spectra observed in the numerical simulations as presented in Fig.~1(g). When the amplitude of surface electron oscillation is at the wave-breaking limit, the harmonics generated via ROW mechanism are expected to follow a power-law relation $I_n\sim n^{-3.5}$ \cite{Yi2021}. This matches the cone channel case, but the cylindrical waveguide produces a slightly softer spectrum, especially at high harmonic numbers. For example, the 2nd and the 8th harmonic produced by the cone channel is roughly $2.7$ and $6.0$ times stronger than that by a cylindrical waveguide, respectively. This is because the surface wave amplitude decreases with time in a cylindrical waveguide, the harmonics produced at later times has a softer spectrum.
Note that a flatter spectrum also means the harmonic generation in the VUV regime ($<200\ {\rm nm}$)  is more efficient in addition to the overall energy boost. For the present cases, the energy of VUV optical vortices reaches 6 mJ (overall efficiency $\sim1.1\%$) in a cone channel, enhanced by more then one order of magnitude than that in a cylindrical waveguide.\\

\begin{figure}[b]
\centering
\subfigure{\includegraphics[width=0.3\textwidth]{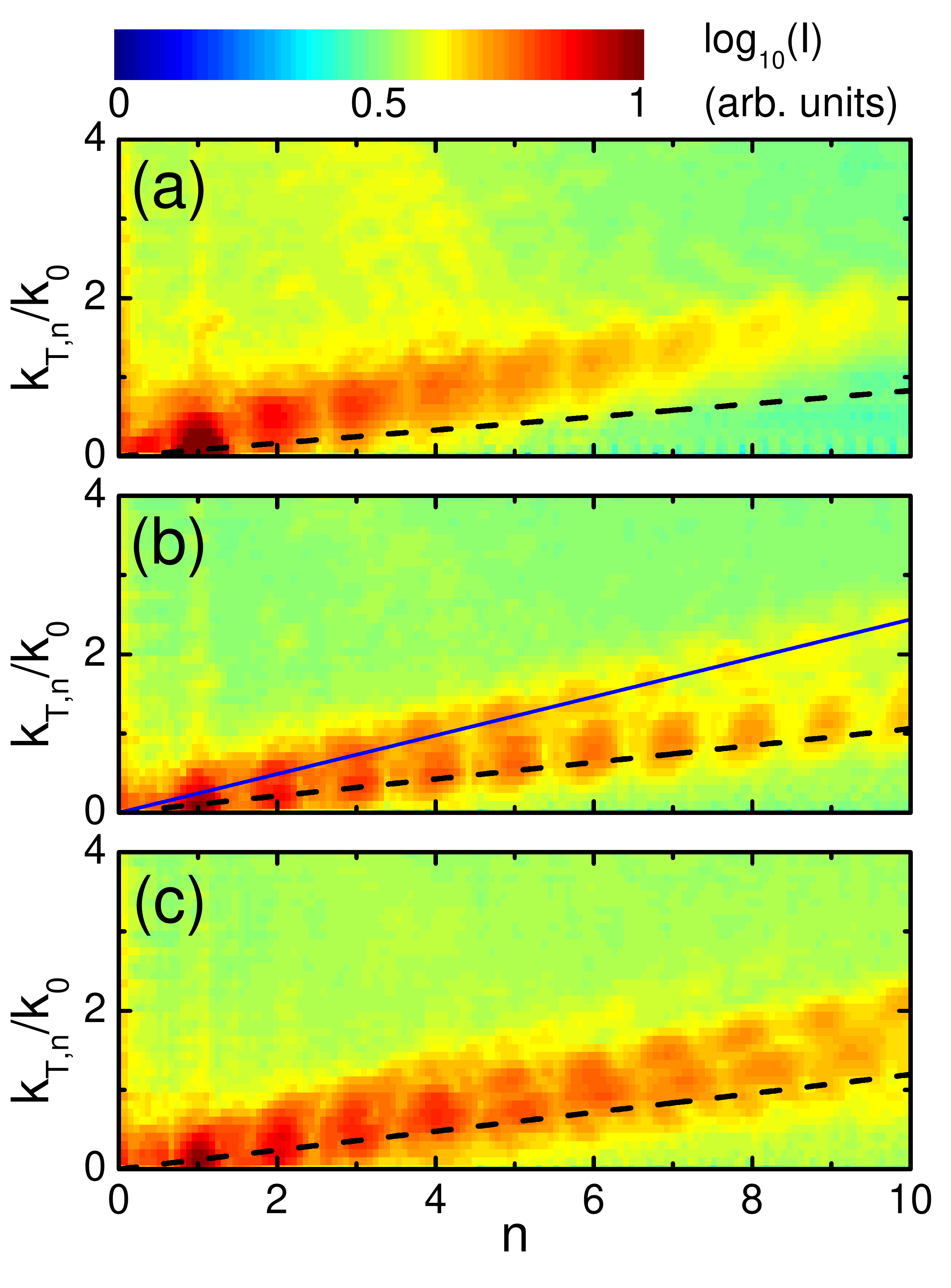}}
%\subfigure{\includegraphics[width=0.3\textwidth]{fig2d}}
\caption{The Intensity distribution of the $E_y$ fields in Fourier space ($\omega - k_{T}$) at propagation distance (a) $L_x=20\ \mu{\rm m}$, (b) $L_x=70\ \mu{\rm m}$, and (c) $L_x=90\ \mu{\rm m}$ in a cone target with $r_1=5\ \mu{\rm m}$ and $r_2=3\ \mu{\rm m}$.
%, at $L_x=70\ \mu{\rm m}$ (d) in a cylindrical target with $r_2=5\ \mu{\rm m}$.
The black dashed line is obtained by $k_{T,n}=n x_1 \lambda_0/2 \pi r$ in each figure, where $r=r_1-(L-L_x)\cdot {\rm tan} \alpha$ is the local radius.
The blue line in (b) marks a second branch of harmonic beams that appears at later times.
    }
\end{figure}

In order to produce high-order harmonics in a long channel, the fundamental and the harmonic light must have the same phase velocity, so that harmonic beams produced at different longitudinal positions can add coherently \cite{Hu2022}, namely $n\omega_0/k_{n,x} = const.$, or equivalently $k_{n,T} \propto n$, where $k_{n,x}$ and $k_{n,T}$ are the longitudinal and transverse wavenumber of the $n$th harmonic, respectively. This is verified by PIC simulations as shown in Fig.~2, where the intensity distribution of the electromagnetic wave in $n-k_{n,T}$ space are obtained via 3D Fourier analyses of the $E_y$ field at different propagation distance. One can see that as the laser travels into the channel, the phase velocities of all the harmonic beams gradually converge, aligning with the driving laser's phase velocity, which can be obtained by plasma waveguide theory \cite{Shen1991,Yi20161}, as $v_p/c\approx 1+k_T^2/(2k_0^2)$, where $k_T = x_1/r$ is the transverse wavenumber of the fundamental mode, and $x_1\approx2.4$ is the first root of eigenvalue equation \cite{Yi20162}.
Note that although $v_p$ depends on the radius $r$, since the cone angle $\alpha$ is very small, the phase velocity variation along a cone channel is negligible, as shown by the black dashed curves in Fig.~2(a-c).

It is worth mentioning that a second branch of harmonic beams appears at around $L_x = 70\ \mu \rm m$, indicated by the blue line in Fig.~2(b).
This coincides with the second phase of rapid increasing of harmonic energy as shown in Fig.~1(e). Such intermittent HHG phases appear because the diffraction effect at the entrance causes the drive laser to bounce between channel walls with large angles before it is coupled fully into waveguide modes. During this phase, the laser intensity on the channel wall varies periodically, high-order harmonics are mostly generated when the intensity is at maximum, and typically with a large angle with respect to the channel axis (Fig.~2(a)), which corresponding to a higher phase velocity.  When the propagation distance is sufficiently long, the laser is fully coupled into waveguide mode, the high-order harmonics are generated continuously, and the phase velocities of the harmonics also decrease to match with the driver, as shown by Fig.~2(b-c). We note that such diffraction effect only occurs near the entrance, it complicates the interaction between laser and plasma channel, as the drive laser beam can not be described as stable waveguide modes, but it doesn't change the underlying physics of the presented work. The simulation results shown in Fig.~2 indicate that the phase matching is fulfilled at all times.\\

\begin{figure}[t]
\centering
\includegraphics[width=0.48\textwidth]{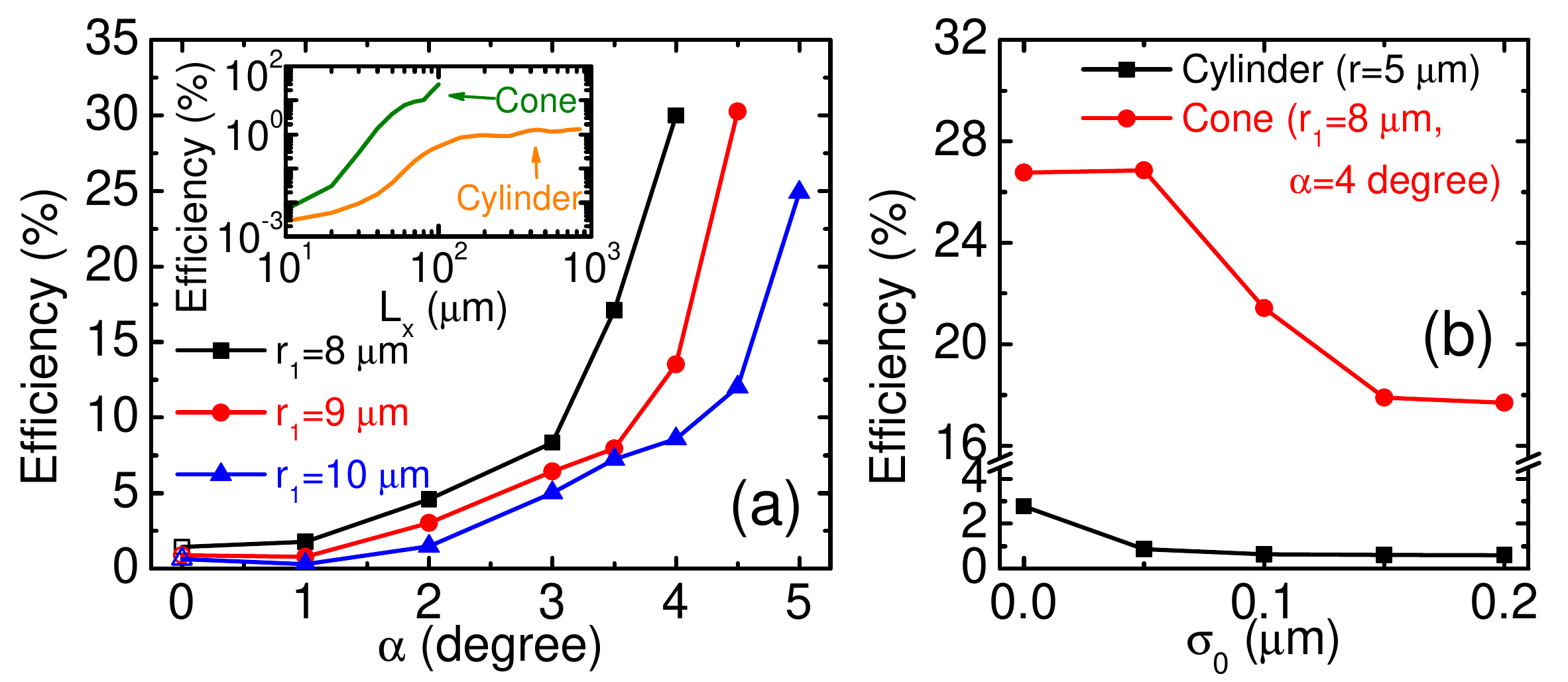}
\caption{
(a) HHG efficiency as a function of cone angle $\alpha$ for different $r_1$. The length $L=100\ \mu \rm m$ is fixed and $r_2$ is adjusted accordingly.
The open markers at $\alpha=0^{\circ}$ refers to the saturated efficiencies obtained in a sufficiently long cylindrical waveguides.
The inset shows the efficiency versus laser propagation distance for a hollow cone ($r_1=8\ {\mu \rm m}, \alpha=4^{\circ}, L=100\ {\mu \rm m}$) and a cylindrical waveguide ($r_1=8\ {\mu \rm m}, \alpha=0^{\circ}, L=1000\ {\mu \rm m}$).
(b) Comparison of HHG efficiency at different preplasma scale lengths between a cone target ($r_1 = 8~\mu m$, and $\alpha = 4^\circ$) and a cylindrical waveguide (radius $r = 5~\mu m$).
    }
\end{figure}

In the following, we examine the parametric dependence of the proposed scheme. It should firstly be noted that similar to the cylindrical waveguide case, the waist of laser focal spot $w_0$ should be smaller than the channel radius at the entrance to avoid producing to many hot electrons via direct impact of laser beam and the edge of the channel \cite{Hu2022}. So in the discussion below we fix the laser waist to be half of the entrance radius $w_0 = r_1/2$.

A cone target introduces a new degree of freedom, namely the cone angle, that can be utilized to control the properties of the HHG process. In Fig. 3(a), harmonics generated by cylindrical and conical targets with varied radii and angles are compared. The parameter scan is performed with 2D PIC simulations with higher resolutions ($dx \times dy=\lambda_0/300 \times \lambda_0/150$).
The length of the cone channels are fixed to be $L = 100\ \mu m$ for the cases presented in Fig.~3(a), and the maximum cone angles are chosen to ensure the channel is not closed at the exit, i.e. the radius of the exit to be $r_2>1\ \mu m$.
The simulation results suggest the overall efficiencies for HHG become higher with increasingly larger cone angles, and smaller radius typically results in a higher HHG efficiency at the same cone angle. As one can see, the overall harmonic efficiencies are over $20\%$ for all the radii cases under consideration, given a sufficiently large cone angle is applied. This is promising for experimental realization of the proposed scheme, as a large entrance radius relaxes the requirement for laser alignment significantly.
The open markers at $\alpha = 0$ in Fig.~3(a) shows the HHG efficiency obtained in a sufficiently long cylindrical waveguide. A typical example (with $r_1 = 8\ \mu m$) is presented in the inset. Obviously, the HHG process is enhanced significantly in a cone target, the HHG efficiency saturates at $\sim 1.4\%$ when the propagation distance approaches $\sim400\ \mu{\rm m}$ in a cylindrical waveguide, about $20$ times lower than that generated in a cone channel with $\alpha = 4^\circ$ and $L \sim 100\ \mu m$.

Finally, we consider the effect of laser prepulse that leads to the expansion of the plasma channel by 3D PIC simulations .
The resulted density ramp on the inner surface can be modeled with $n(r<r_0)=n_0{\rm exp}[(r-r_0)/\sigma_0]$, where $\sigma_0$ is the scale length. In Fig.~3(b) we plot the HHG efficiencies for conical and cylindrical channels as functions of the preplasma scale length. Importantly, the efficiency decreases in all case when the plasma expansion is significant.
When a preplasma is present, the electron layers at different radial
locations tend to oscillate at different phases. The transverse ponderomotive force of the laser tends to compress the preplasma, but this becomes increasingly challenging as the preplasma scale length growth due to laser heating \cite{Li2019}.
As a result, a preplasma tends to weaken the compression of the reflection electrons and hence reduce coherence of local HHG.
Thus, high-contrast high-power laser is required for such application, which can be achieved using plasma mirrors  \cite{Thaury2007,Kahaly2013} and cross-polarized wave generation technique \cite{Jullien2006}.

Interestingly, the numerical results also suggest that hollow-cone targets could mitigate this effect for reasonably small preplasma scale length. When $\sigma_0=0.05\ \mu{\rm m}$, the overall efficiency declined by $70\%$ for the cylindrical channel compared with the case of no preplasma, but for a cone target, the efficiency is almost the same. When $\sigma_0=0.1\ \mu{\rm m}$ and $0.2\ \mu{\rm m}$, the HHG efficiency for the conical channels are $21\%$ and $14\%$, respectively. In both cases it is improved by roughly 30 times comparing with in a cylindrical waveguide.
This could also be attributed to the focusing effect of the laser in a conical channel, as intensity grows, the transverse ponderomotive force tends to compress the pre-plasma thus leads to compact oscillations of surface electrons. A similar idea is recently implemented experimentally using the ROM mechanism, where the preplasma is compressed by a CP laser beam before the main pulse arrives, in order to enhance the harmonic beam intensities \cite{Li2022}.

In Summary, we propose and numerically demonstrate an efficient scheme to generate tens-of-mJ harmonic vortices, from a hollow cone target irradiated by a joule-level femtosecond driving laser pulse.
The driving laser tends to be focused and intensified as it propagates in the cone target,
which helps to maintain a high-amplitude surface wave for a long distance.
As a result, the efficiency for high-order harmonic vortices generation is found to be nearly one order of magnitude higher than that form normal cylindrical targets, reaching $>20\%$.
The scheme potentially paves the way to the development of powerful vortices sources in near future.

\begin{acknowledgments}
This work is supported by the National Key R$\&$D Program of China (No. 2021YFA1601700), the Shanghai Pujiang Talent Plan (No. 21PJ1407500), and the National Natural Science Foundation of China (No. 12205187).
\end{acknowledgments}

%\begin{acknowledgments}
%This work was supported by the Thousand Youth Talents Plan...
%\end{acknowledgments}

%\section*{aip publishing data sharing policy}
%The data that support the findings of this study are available from the corresponding author upon reasonable request.

\end{document}